\documentclass[aps,prb,twocolumn,showpacs,superscriptaddress,groupedaddress]{revtex4}  
\usepackage{graphicx, epstopdf}  
\usepackage{dcolumn}   
\usepackage{bm}        
\usepackage{amssymb,amsmath}   
\usepackage{color,epsfig,float}

\begin{document}


\title{Transport properties of a two-dimensional electron gas dressed by light}
\author{S. Morina}
\affiliation{Division of Physics and Applied Physics, Nanyang
Technological University 637371,  Singapore} \affiliation{Science
Institute, University of Iceland, Dunhagi 3, IS-107, Reykjavik,
Iceland}

\author{O.V. Kibis}\email{Oleg.Kibis(c)nstu.ru}
\affiliation{Department of Applied and Theoretical Physics, Novosibirsk State Technical University,\\
Karl Marx Avenue 20, Novosibirsk 630073, Russia}
\affiliation{Division of Physics and Applied Physics, Nanyang
Technological University 637371,  Singapore}

\author{A.A. Pervishko}
\affiliation{Division of Physics and Applied Physics, Nanyang
Technological University 637371,  Singapore}

\author{I.A. Shelykh}
\affiliation{Division of Physics and Applied Physics, Nanyang
Technological University 637371,  Singapore} \affiliation{Science
Institute, University of Iceland, Dunhagi 3, IS-107, Reykjavik,
Iceland}\affiliation{ITMO University, St. Petersburg 197101,
Russia}


\begin{abstract}
We show theoretically that the strong interaction of a
two-dimensional electron gas (2DEG) with a dressing
electromagnetic field drastically changes its transport
properties. Particularly, the dressing field leads to the giant
increase of conductivity (which can reach thousands of percents),
results in nontrivial oscillating dependence of conductivity on
the field intensity, and suppresses the weak localization of 2DEG.
As a consequence, the developed theory opens an unexplored way to
control transport properties of 2DEG by a strong high-frequency
electromagnetic field. From experimental viewpoint, this theory is
applicable directly to quantum wells exposed to a laser-generated
electromagnetic wave.
\end{abstract}

 \maketitle

\section{Introduction}
Transport properties of two-dimensional electron gas (2DEG) in
nanostructures exposed to a high-frequency electromagnetic field
have been studied in the deep past and taken deserved place in
textbooks (see, e.g.,
Refs.~[\onlinecite{FerryGoodnick,Davies,Nag}]). However, the most
attention in previous studies on the subject was paid to the
regime of weak light-matter interaction. Following the
conventional terminology of quantum optics, in this regime an
electron energy spectrum is assumed to be unperturbed by photons.
Correspondingly, the weak electron-photon interaction results only
in electron transitions between unperturbed electron states, which
are accompanied by absorption and emission of photons. As a
consequence, the regime of weak electron-photon interaction in
solids leads to photovoltaic effects, high-frequency conductivity
and other well-known electronic transport phenomena which are
accompanied with absorption of field energy by conduction
electrons. However, the interaction between electrons and a strong
electromagnetic field (the regime of strong light-matter
interaction) cannot be described as a weak perturbation. In this
case, the system ``electron + electromagnetic field'' should be
considered as a whole. Such a bound electron-photon system, which
was called ``electron dressed by photons'' (dressed electron),
became a commonly used model in modern physics.
\cite{Scully,Cohen-Tannoudji} For a long time, the main objects
for studying the physical properties of dressed electrons were
atoms and molecules. The field-induced modification of the energy
spectrum and wave functions of dressed electrons --- the so-called
dynamic Stark effect --- was discovered in atoms many years ago
\cite{Autler_55} and has been studied in details in various atomic
and molecular systems. These studies of strong electron-photon
processes formed up such an exciting field of modern physics as
quantum optics. \cite{Scully,Cohen-Tannoudji} In nanostructures,
the research activity in the area of quantum optics was focused on
exciton-polaritonic effects in microcavities with quantum wells
\cite{Kasprzak,Balili,Yamamoto,KenaCohen,Plumhof,Mazzeo,Smolka}
and quantum dots,
\cite{QDStrongCouplingPillar,QDStrongCouplingDefect,Johne}
physical properties of dressed electrons in graphene
\cite{Kibis_10,Savelev2011,Kibis_2011} and quantum wires,
\cite{Kibis_11} variety of technological applications,
{\cite{PolDevices}} including novel types of the lasers,
\cite{Christopoulos,Schneider} optical switches and logic gates,
\cite{Paraiso,Amo} all-optical integrated circuits \cite{Espinosa}
and others.  As to transport properties of dressed 2DEG, they are
still waiting for detailed research. The present article is aimed
to fill partially this interdisciplinary gap which takes place at
the border between physics of nanostructures and quantum optics.

\section{Model} 
\begin{figure}[h]
\includegraphics[width=\linewidth]{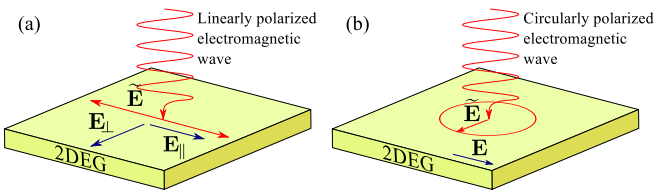}
\caption{(Color online) Sketch of the system under consideration:
The 2DEG dressed by (a) linearly polarized high-frequency field
$\widetilde{\mathbf{E}}$ and (b) circularly polarized one in the
presence of a stationary in-plane electric field
$\mathbf{E}=(E_\parallel,{E}_\perp)$.} \label{setup}
\end{figure}

For definiteness, we will restrict our
consideration to the case of 2DEG with a parabolic electron energy
spectrum
\begin{equation}\label{energy}
\varepsilon_\mathbf k=\frac{\hbar^2\mathbf k^2}{2m},
\end{equation}
where $\mathbf{k}$ is the electron wave vector, and $m$ is the
electron effective mass. Let the 2DEG be subjected to a plane
monochromatic electromagnetic wave propagating perpendicularly to
the 2DEG plane (see Fig.~\ref{setup}) . In what follows, we will
assume that the wave frequency, $\omega$, meets two conditions.
Firstly, the wave frequency is far from resonant electron
frequencies corresponding to interband electron transitions and,
therefore, the interband absorption of the wave by the 2DEG is
absent. Secondly, the wave frequency is high enough in order to
satisfy the inequality
\begin{equation}\label{omega}
\omega\tau_0\gg1,
\end{equation}
where $\tau_0$ is the electron relaxation time in an unirradiated
2DEG. It is well-known that the intraband (collisional) absorption
of wave energy by conduction electrons is negligibly small under
condition (\ref{omega}) (see, e.g.,
Refs.~[\onlinecite{Ashcroft,Harrison,Kibis_14}]). Thus, the
considered electromagnetic wave can be treated as a purely
dressing (nonabsorbable) field. It follows from the basic
principles of quantum optics that the strong coupling of electrons
to such a dressing field leads to the renormalization of all
physical quantities describing the electrons.
\cite{Scully,Cohen-Tannoudji} Particularly, it is well-known that
a high-frequency electromagnetic field can strongly affect the
scattering of conduction electrons and change electronic transport
characteristics. \cite{Melnikov_69,Fomin_88} Recently, this
approach was extended for the case of a two-dimensional electron
gas subjected to a purely dressing field which cannot be absorbed
and emitted by conduction electrons. \cite{Kibis_14} Namely, the
scattering probability of a dressed electron between electron
states with wave vectors $\mathbf{k}$ and $\mathbf{k}^\prime$ per
unit time has the form \cite{Kibis_14}
\begin{equation}\label{W0}
w_{\mathbf{k}^\prime\mathbf{k}}=\frac{2\pi}{\hbar}J_0^2\left(f_{\mathbf{k}^\prime\mathbf{k}}\right)
\left|U_{\mathbf{k}^\prime\mathbf{k}}\right|^2
\delta(\varepsilon_{k^\prime}-\varepsilon_k),
\end{equation}
where $J_0(z)$ is the zeroth order Bessel function of the first
kind, and $U_{\mathbf{k}^\prime\mathbf{k}}$ is the matrix element
of the scattering potential, $U(\mathbf{r})$, which arises from
macroscopically large number of scatterers in the conductor. In
the case of linearly polarized dressing field, \cite{Kibis_14} the
argument of the Bessel function is given by
\begin{equation}\label{flin}
f_{\mathbf{k}^\prime\mathbf{k}}=\frac{e{\mathbf{E}_\omega}
(\mathbf{k}-\mathbf{k}^\prime)}{m\omega^2},
\end{equation}
where ${E_\omega}$ is the amplitude of the electric field of the
wave, $\widetilde{\mathbf{E}}(t)=\mathbf{E}_\omega\sin\omega t$.
In the case of circularly polarized dressing field (see Appendix
A), this argument is
\begin{equation}\label{fcir}
f_{\mathbf{k}^\prime\mathbf{k}} =
\frac{2{E_\omega}ek}{m\omega^2}\sin\left(\frac{
\theta_{\mathbf{k}^\prime\mathbf{k}}}{2}\right),
\end{equation}
where
$\theta_{\mathbf{k}^\prime\mathbf{k}}=(\widehat{\mathbf{k}^\prime,\mathbf{k}})$
is the angle between electron wave vectors $\mathbf{k}$ and
$\mathbf{k'}$. The formal difference between the scattering
probability of dressed electron (\ref{W0}) and the conventional
expression for the scattering probability of bare electron
\cite{Landau_3} consists in the Bessel-function factor
$J_0^2\left(f_{\mathbf{k}^\prime\mathbf{k}}\right)$, where
$f_{\mathbf{k}^\prime\mathbf{k}}$ depends on the dressing field
amplitude ${E_\omega}$ and the dressing field frequency $\omega$
[see Eqs.~(\ref{flin})--(\ref{fcir})]. Just this factor results in
nontrivial dependence of electronic transport properties on the
dressing field. In what follows, we will focus our attention on
the conductivity and the weak localization of dressed 2DEG.

\section{Conductivity of dressed 2DEG} Assuming the temperature
to be zero, let us apply a stationary (dc) electric field
$\mathbf{E}$ to the 2DEG. It follows from the conventional
Boltzmann equation for conduction electrons (see, e.g.,
Ref.~[\onlinecite{Anselm}]) that electric current density,
$\mathbf{J}$, is given by the well-known expression
\begin{equation}\label{j}
\mathbf J =
\frac{e^2}{2\pi^2}\int\limits_\mathbf{k}\left[\mathbf{E\cdot
v(k)}\right]\tau(\mathbf k)\mathbf v(\mathbf
k)\delta(\varepsilon_k-\varepsilon_F)d^2\mathbf k,
\end{equation}
where $v(\mathbf k)=({1}/{\hbar})\nabla_\mathbf
k\varepsilon_\mathbf k$ is the electron velocity,
$\varepsilon_F=\hbar^2\mathbf{k}^2_F/2m$ is the electron Fermi
energy, and $\tau(\mathbf k)$ is the relaxation time. In the most
general case of anisotropic electron scattering, this relaxation
time is given by the equation \cite{Sorbello}
\begin{equation}\label{tau}
\frac{1}{\tau(\mathbf k)} = \sum_{\mathbf k'}{\left[1 -
\frac{\tau(\mathbf k')\mathbf{E\cdot v(k')}}{\tau(\mathbf
k)\mathbf{E\cdot v(k)}}\right] w_\mathbf{k'k}}.
\end{equation}
Since absorption of a high-frequency field satisfying inequality
(\ref{omega}) is negligibly small, the field does not change the
equilibrium distribution function of electrons. \cite{Kibis_14}
Therefore, the field influences on the stationary (dc) transport
of dressed electrons only through the renormalization of the
scattering probability, $w_{\mathbf{k}^\prime\mathbf{k}}$, given
by Eq.~(\ref{W0}). It should be stressed that the condition
(\ref{omega}) is crucial in order to consider a 2DEG subjected to
the field as an equilibrium system. Otherwise, the photon-assisted
scattering of electrons is accompanied by absorption of field
energy and leads to heating 2DEG. In this case, there are no
stationary transport characteristics of the 2DEG and the problem
should be reformulated in terms of the non-equilibrium electron
transport.

Substituting the scattering probability of dressed electron
(\ref{W0}) into Eq.~(\ref{tau}), we can obtain from
Eqs.~(\ref{j})--(\ref{tau}) the conductivity of dressed 2DEG,
$\sigma_{ij}=J_i/E_j$.

To simplify calculations, let us consider the electron scattering
within the $s$-wave approximation, \cite{Landau_3} where the
matrix elements $U_{\mathbf{k}^\prime\mathbf{k}}$ do not depend on
the angle
$\theta_{\mathbf{k}^\prime\mathbf{k}}=(\widehat{\mathbf{k}^\prime,\mathbf{k}})$.
Substituting Eq.~(\ref{fcir}) into Eqs.~(\ref{W0}) and
(\ref{j})--(\ref{tau}), we arrive at the isotropic conductivity of
2DEG dressed by circularly polarized field, $\sigma_c$, which is
given by the expression
\begin{equation}\label{sc}
\frac{\sigma_\mathrm{c}}{\sigma_0} = 2\pi\left[\int_0^{2\pi}
[1-\cos\theta]J_0^2\left(\frac{2eE_\omega k_F}{m\omega^2}\sin\frac
\theta 2\right)d\theta\right]^{-1},
\end{equation}
where $\sigma_0$ is the conductivity of an unirradiated 2DEG. In
the case of linearly polarized dressing field, the conductivity
tensor $\sigma_{ij}$ has a simple diagonal form,
$\sigma_{ii}=(\sigma_\parallel,\sigma_\perp)$, in the basis of two
axes which are parallel ($\parallel$) and perpendicular ($\perp$)
to the wave field $\widetilde{\mathbf E}$, respectively.
Substituting Eq.~(\ref{flin}) into Eqs.~(\ref{W0}) and
(\ref{j})--(\ref{tau}), we arrive at
\begin{equation}\label{spar}
\frac{\sigma_\parallel}{\sigma_0} =
\frac{1}{\pi}\int_0^{2\pi}\frac{\tau_\parallel(\theta)}{\tau_0}\cos^2\theta
d\theta
\end{equation}
and
\begin{equation}\label{sperp}
\frac{\sigma_\perp}{\sigma_0} =
\frac{1}{\pi}\int_0^{2\pi}\frac{\tau_\perp(\theta)}{\tau_0}\sin^2\theta
d\theta
\end{equation}
where $\tau_\parallel(\theta)$ and $\tau_\perp(\theta)$ are the
relaxation time (\ref{tau}) at the Fermi energy for the dc
electric field $\mathbf{E}=(E_\parallel,0)$ and
$\mathbf{E}=(0,E_\perp)$, respectively (see Fig.~1a), and
$\theta=(\widehat{\mathbf{k},\mathbf{E}})$.

\begin{figure}[h]
\includegraphics[width=\linewidth]{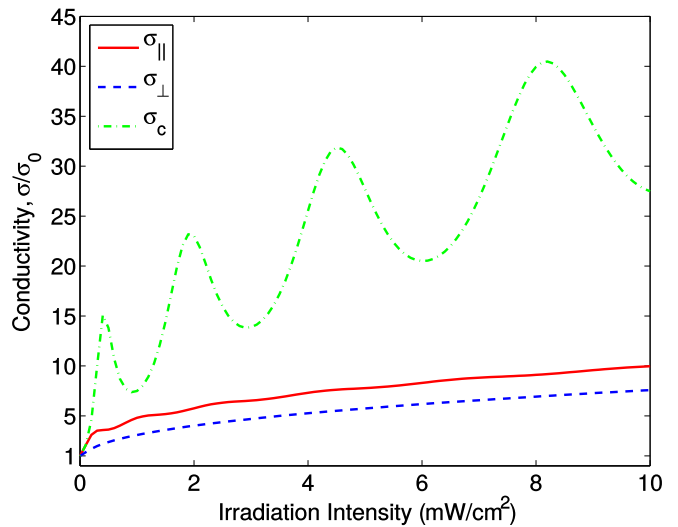}
\caption{(Color online) The conductivity of 2DEG in a GaAs quantum
well irradiated by a dressing electromagnetic field with the
frequency $\omega=10^{11}$~rad/s at the temperature $T=0$ for
$\varepsilon_F=10$~meV.} \label{conductivity}
\end{figure}

The conductivities of irradiated 2DEG (\ref{sc})--(\ref{sperp}),
which can be easily calculated numerically, are plotted in
Fig.~\ref{conductivity}. It is seen that the irradiation of 2DEG
by a dressing light results in giant increase of conductivity,
which can reach thousands of percents. Physically, this increase
follows from the fact that the dressing field significantly
decreases the scattering probability of dressed 2DEG (\ref{W0}).
From the mathematical viewpoint, this is a consequence of the
rapid decrease of the Bessel function in Eq.~(\ref{W0}) versus the
dressing field amplitude $E_\omega$. As to the oscillations of the
plotted conductivity, this is a formal consequence of the
oscillating behavior of the Bessel function in the probability
(\ref{W0}). In order to give a qualitative physical explanation of
this behavior of conductivity, we have to stress that the Born
scattering probability (\ref{W0}) is described by the overlap of
wave functions of an incident electron with the wave vector
$\mathbf{k}$ and a scattered electron with the wave vector
$\mathbf{k}^\prime$ in the area of scattering potential $U$. In
turn, this overlap depends on the difference of electron phases in
the high-frequency field, which are presented by the terms with
$\sin\omega t$ and $\cos\omega t$ in Eq.~(\ref{Apsi}) (see also
Eq.~(6) in Ref.~\onlinecite{Kibis_14}). Since the terms depend on
the field amplitude $E_\omega$, the stationary overlap of the wave
functions
--- and, correspondingly, the scattering probability (\ref{W0})
--- can vanish for certain field amplitudes which correspond to the zeros of the Bessel function. As
a consequence, this leads to both the increasing of conductivity
and the oscillations of the conductivity, which are seen in
Fig.~\ref{conductivity}.

The difference between the conductivities $\sigma_\parallel$ and
$\sigma_\perp$ arises from the scattering anisotropy of a 2DEG
dressed by a linearly polarized field, which follows directly from
Eqs.~(\ref{W0})-(\ref{flin}).  It should be stressed that the
arguments of the Bessel function in the scattering probability
(\ref{W0}) for the cases of linearly polarized field (\ref{flin})
and circularly polarized field (\ref{fcir}) strongly differ from
each other. Therefore, the integration over $\mathbf{k}$ and
$\mathbf{k}^\prime$    and   in the scattering probability (3)
substituted into the kinetic Boltzmann equation leads to strongly
different plots in Fig.~2 for linear polarization and circular
polarization. Particularly, the amplitude of oscillations of
conductivity in the case of linear polarization is nonzero but
very small as compared to the case of circular polarization.
Therefore, the most appropriate experimental set for observing the
oscillations should be based on using circularly polarized light.

\section{Weak localization of dressed 2DEG} Multiple scattering
of electrons in a conductor leads to the self-interference of
electron waves propagating along a closed trajectory in mutually
opposite
---  clockwise and counterclockwise --- directions. As a result of
the interference, the well-known weak localization (WL) of
conduction electrons appears. \cite{Altshuler,
Bergmann,Chalravarty} To find the WL correction to the
conductivity of dressed 2DEG, $\Delta\sigma$, one needs to
consider all possible closed electron trajectories in the 2DEG
irradiated by a dressing electromagnetic field. Going this way,
the sought correction can be calculated using the conventional
expression {\cite{Altshuler}}
\begin{equation}\label{sWL}
\frac{\Delta
\sigma}{\sigma_0}=-\frac{h}{m}\int_{\tau_{F}}^{{\tau}_{\varphi}}
C(t) dt,
\end{equation}
where $ C(t)$ is the probability to find an electron in the
initial point of its trajectory at the time $t$ averaged over 2DEG
plane, $\tau_{F}$ is the mean free time of conduction electron at
the Fermi energy, which is given by the expression
\begin{equation}\label{tF}
\frac{1}{\tau_{F}} =
\sum_{\mathbf{k}^\prime_F}w_{\mathbf{k}_F\mathbf{k}^\prime_F},
\end{equation}
and $\tau_\varphi$ is the effective breaking-time of electron
phase. Generally, electron phase can be broken by both inelastic
scattering and a magnetic field which destroy WL.
{\cite{Altshuler1,Aronov, Hassenkam, Pedersen, Altshuler2,
Zduniak}} The breaking-time of electron phase caused by a magnetic
field, $B$, is $\tau_{\varphi B}={\hbar}/{4eDB}$, where $D$ is the
diffusion constant. {\cite{Zduniak}} Correspondingly, the
effective breaking-time of electron phase, $\tau_\varphi$, can be
written as
$1/\tau_\varphi=1/\tau_{\varphi\varepsilon}+1/\tau_{\varphi_B}$,
where $\tau_{\varphi\varepsilon}$ is the characteristic
breaking-time arisen from inelastic processes. As to the mean free
time of conduction electron, $\tau_{F}$, it can be found by
substituting the scattering probability of dressed electron
(\ref{W0}) into Eq.~(\ref{tF}). It has been mentioned above that
this scattering probability is anisotropic for 2DEG dressed by a
linearly polarized field. Therefore, the time (\ref{tF}) depends
on the initial electron wave vector $\mathbf{k}_F$. To describe
this anisotropy, it is suitable to introduce the two mean free
times: $\tau_{{F}\parallel}$ and $\tau_{{F}\perp}$ for the cases
of $\mathbf{k}_F\parallel\widetilde{\mathbf{E}}$ and
$\mathbf{k}_F\perp\widetilde{\mathbf{E}}$, respectively.
Substituting these two times into Eq.~(\ref{sWL}), we arrive at
the two WL corrections, $\Delta\sigma_\parallel$ and
$\Delta\sigma_\perp$. These two corrections describe the
conductivity of dressed 2DEG for the cases of
$\mathbf{E}=(E_\parallel,0)$ and $\mathbf{E}=(0,E_\perp)$,
respectively (see Fig.~1a). Certainly, in the case of circularly
polarized dressing field, the anisotropy vanishes and the
corresponding WL correction, $\Delta\sigma_c$, does not depend on
direction in the 2DEG plane.

Within the conventional theory of WL, the probability $C(t)$ in
Eq.~(\ref{sWL}) can be described by the Feynman path integral
{\cite{Chalravarty, Rammer}}
\begin{eqnarray}\label{C}
C(t)&=&\int\mathcal{D}x\mathcal{D}y
\exp\Bigg[-\int_{-{t}/{2}}^{{t}/{2}} \frac{\dot{x}^2}{4D_{x}}+
\frac{\dot{y}^2}{4D_{y}}\nonumber\\
&-&\frac{i}{\hbar}e\mathbf{r}[\widetilde{\mathbf{E}}(t')-
\widetilde{\mathbf{E}}(-t')]dt'\Bigg],
\end{eqnarray}
where the integration should be performed over closed electron
trajectories per unit plane of 2DEG, and $\mathbf{r}=(x,y)$ is the
electron radius-vector in the 2DEG plane. In the integrand of the
path integral (\ref{C}), the first two terms in the exponent
describe the kinetic energy of the diffusion propagation of a
dressed electron along the path, where $D_{x,y}$ are the diffusion
constants along the $x,y$ axes. Generally, these constants can be
written as $D_{x}=v_F^2\tau_{F\parallel}/2$ and
$D_{y}=v_F^2\tau_{F\perp}/2$ (see, e.g.,
Ref.~[\onlinecite{Zduniak}]), where $v_F$ is the Fermi velocity,
and the $x$ axis is assumed to be directed along the electric
field $\widetilde{\mathbf{E}}$ of linearly polarized dressing
field. In the case of circularly polarized dressing field, we have
$\tau_{F\parallel}=\tau_{F\perp}$ and, therefore, the $x,y$ axes
can be chosen arbitrary. The last term in the exponent takes into
account the potential energy of an electron in the dressing field
and describes the self-interference of the dressed electron
between the time-reversed trajectories. Performing the path
integration in Eq.~(\ref{C}) within the conventional procedure
\cite{Feynman} and substituting the obtained probability (\ref{C})
into Eq.~(\ref{sWL}), we arrive at the WL corrections to the
conductivity of dressed 2DEG, which are presented in Fig.~3.

It is seen in Fig.~3 that the irradiation of 2DEG by a dressing
field leads to decreasing WL corrections to the conductivity.
Physically, this is a consequence of suppressed scattering in a
dressed 2DEG, which has been mentioned above. As a result of the
suppression, the probability of electron movement along a closed
trajectory in the dressed 2DEG decreases and, correspondingly, WL
corrections to the conductivity of dressed 2DEG decrease as well.
It should be stressed that this effect differs significantly from
the known suppression of WL due to intraband absorption of
irradiation by conduction electrons.
\cite{Altshuler1981,Wang1987,Altshuler1998} Indeed, under the
condition (\ref{omega}) there is no absorption of high-frequency
field by conduction electrons. Therefore, the destruction of WL,
which arises from the breaking of electron phase in inelastic
electron-photon processes,
\cite{Altshuler1981,Wang1987,Altshuler1998} is negligibly small in
the 2DEG under consideration. As a consequence of decreasing WL
effects, the influence of a magnetic field on WL corrections to
the conductivity for a dressed 2DEG is less than one for a bare
2DEG (see the insert in Fig.~3). It should be noted that the WL
correction to the conductivity of 2DEG has a logarithmic
dependence on a magnetic field. \cite{Altshuler} Although the
irradiation on the 2DEG reduces WL corrections, the behavior of
conductivity of dressed 2DEG in a magnetic field obeys the same
law. As to small oscillations visible in Fig.~3, they have the
same nature as oscillations of conductivity pictured in Fig.~2.
Anisotropy of weak localization in the case of linearly polarized
field is within one percent and cannot be easily detected
experimentally. Therefore, we have
$\Delta\sigma_\parallel\approx\Delta\sigma_\perp$ for the plots in
Fig.~3.

\begin{figure}[h]
\centering
\includegraphics[width=\linewidth]{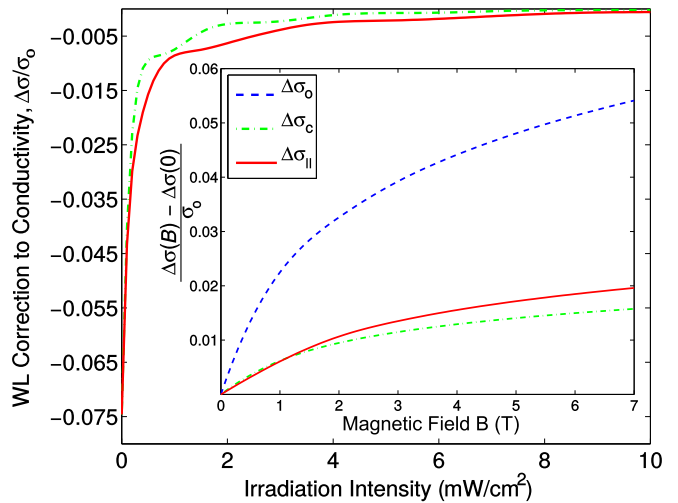}
\caption{(Color online) The dependence of weak localization
corrections to conductivity of 2DEG on intensity of irradiation by
a dressing field. Calculations are performed for 2DEG in a GaAs
quantum well at the temperature $T=0$ for $\varepsilon_F=10$~meV
and $\tau_{\varphi\varepsilon}=80$~ps. The insert demonstrates the
dependence of the corrections on magnetic field directed
perpendicularly to the 2DEG in the presence of the irradiation
with the intensity $I=3$~mW/cm$^2$.} \label{localizationeffects}
\end{figure}

\section{Conclusion} We demonstrated theoretically that the strong
coupling of 2DEG to a dressing electromagnetic field results in a
set of such unexpected transport phenomena as a giant increase of
conductivity, oscillating behavior of conductivity versus
intensity of dressing field and suppression of weak localization
effects. These phenomena open a new way to control the transport
properties of 2DEG by a strong high-frequency electromagnetic
field and, therefore, form a basis for novel optoelectronic
nanodevices.

\begin{acknowledgments} The work was partially supported by FP7
IRSES projects POLATER and QOCaN, FP7 ITN project NOTEDEV, Tier 1
project ``Polaritons for novel device applications'', Rannis
project BOFEHYSS,  RFBR project 14-02-00033, and the Russian
Ministry of Education and Science.
\end{acknowledgments}

\appendix
\section{Scattering of 2DEG dressed by a circularly polarized electromagnetic field}
For definiteness, we will restrict our consideration to the case
of a two-dimensional electron system subjected to a circularly
polarized electromagnetic wave propagating perpendicularly to the
system. Let the system lie in the plane $(x,y)$ at $z=0$ and the
wave propagate along the $z$ axis. Then the electron properties of
the system in the absence of scatterers are described by the
Schr\"odinger equation
\begin{equation}\label{Ashr}
i\hbar\frac{\partial\psi}{\partial t}=\hat{\cal{H}}_0\psi
\end{equation}
with the Hamiltonian
$\hat{\cal{H}}_0=[\hat{\mathbf{p}}-e\mathbf{A}(t)]^2/2m$, where
$\hat{\mathbf{p}}=(\hat{{p}}_x,\hat{{p}}_y)$ is the operator of
electron momentum, $m$ is the effective electron mass, $e$ is the
electron charge,
$$\mathbf{A}(t)=\mathbf{e}_x(E_\omega/\omega)\sin\omega
t+\mathbf{e}_y(E_\omega/\omega)\cos\omega t$$ is the vector
potential of the wave, $E_\omega$ is the electric field amplitude
of the wave, $\omega$ is the frequency of the wave, and
$\mathbf{e}_{x,y}$ are the unit vectors along the $x,y$ axis. The
Schr\"odinger equation (\ref{Ashr}) can be solved accurately and
the exact wave function of the electron is
\begin{eqnarray}\label{Apsi}
\psi_\mathbf{k}(\mathbf{r},t)&=&
\exp\left[-i\left(\frac{\varepsilon_k
t}{\hbar}+\frac{E_\omega^2e^2t}{2m\omega^2\hbar}-\frac{E_\omega
ek_y}{m\omega^2}\sin\omega
t\right.\right.\nonumber\\
&+&\left.\left.\frac{E_\omega ek_x}{m\omega^2}\cos\omega
t\right)\right]\varphi_{\mathbf{k}}(\mathbf{r}),
\end{eqnarray}
where
$\varphi_{\mathbf{k}}(\mathbf{r})={V}^{-1/2}\exp(i\mathbf{k}\mathbf{r})$
is the plane electron wave, $\mathbf{k}=(k_x,k_y)$ is the electron
wave vector, $\mathbf{r}=(x,y)$ is the electron radius-vector, $V$
is the normalization volume, and $\varepsilon_k={\hbar^2k^2}/{2m}$
is the energy spectrum of a free electron. Evidently, the wave
function (\ref{Apsi}) can be easily verified by direct
substitution into the Schr\"odinger equation (\ref{Ashr}).
Introducing the polar system $\{k,\varphi\}$, we can write the
electron wave vectors as $k_x=k\cos\varphi$ and
$k_y=k\sin\varphi$. Then the wave function (\ref{Apsi}) takes the
simplest form
\begin{eqnarray}\label{Apsis}
\psi_\mathbf{k}(\mathbf{r},t)&=&\exp\left[-i\left(\frac{\varepsilon_k
t}{\hbar}+\frac{E_\omega^2e^2t}{2m\omega^2\hbar}\right.\right.\nonumber\\
&+&\left.\left.\frac{E_\omega ek}{m\omega^2}\cos(\omega
t+\varphi)\right)\right]\varphi_{\mathbf{k}}(\mathbf{r}).
\end{eqnarray}

Let an electron move in a scattering potential $U(\mathbf{r})$ in
the presence of the same field. Then the wave function of the
electron, $\Psi(\mathbf{r},t)$, satisfies the Schr\"odinger
equation
\begin{equation}\label{ASE}
i\hbar\frac{\partial\Psi(\mathbf{r},t)}{\partial t}=[\hat{\cal
H}_0+U(\mathbf{r})]\Psi(\mathbf{r},t).
\end{equation}
Assuming the scattering potential energy $U(\mathbf{r})$ to be a
small perturbation, we can apply the conventional perturbation
theory to describe the electron scattering. Since the functions
(\ref{Apsis}) with different wave vectors $\mathbf{k}$ form the
complete function system for any time $t$, we can seek solutions
of the Schr\"odinger equation (\ref{ASE}) as an expansion
\begin{equation}\label{AP}
\Psi(\mathbf{r},t)=\sum_{\mathbf{k}^\prime}a_{\mathbf{k}^\prime}(t)\psi_{\mathbf{k}^\prime}(\mathbf{r},t).
\end{equation}
Let an electron be in the state (\ref{Apsis}) with the wave vector
$\mathbf{k}$ at the time $t=0$. Correspondingly,
$a_{\mathbf{k}^\prime}(0)=\delta_{\mathbf{k}^\prime,\mathbf{k}}$,
where $\delta_{\mathbf{k}^\prime,\mathbf{k}}$ is the Kronecker
symbol. In what follows, we will assume that the wave frequency
$\omega$ is large enough to satisfy the inequality
\begin{equation}\label{Atau}
\omega\tau_0\gg1,
\end{equation}
where $\tau_0$ is the characteristic relaxation time of conduction
electron in the absence of the wave. Under the condition
(\ref{Atau}), we can neglect the absorption (emission) of field
energy by a scattered electron (see, e.g.,
Refs.~[\onlinecite{Ashcroft,Harrison}]). Within this
approximation, the scattering potential $U(\mathbf{r})$ mixes only
electron states $\mathbf{k}$ and $\mathbf{k}^\prime$ with the same
energy, $\varepsilon_k=\varepsilon_{k^\prime}$. \cite{Kibis_14}
Therefore, it is enough to take into account only terms with
$k^\prime=k$ in the expansion (\ref{AP}). Substituting the
expansion (\ref{AP}) into the Schr\"odinger equation (\ref{ASE})
and restricting the accuracy by the first order of perturbation
theory, we arrive at the expression
\begin{equation}\label{Aak}
a_{\mathbf{k}^\prime}(t)=-i\frac{U_{\mathbf{k}^\prime\mathbf{k}}}{\hbar}
\int\limits_0^te^{i(\varepsilon_{k^\prime}-\varepsilon_k)t^\prime/{\hbar}}e^{if_{\mathbf{k}^\prime\mathbf{k}}
\sin[\omega t^\prime+(\varphi+\varphi^\prime)/2]}dt^\prime,
\end{equation}
where
\begin{equation}\label{AU}
U_{\mathbf{k}^\prime\mathbf{k}}=\left\langle\varphi_{\mathbf{k}^\prime}(\mathbf{r})
\left|U(\mathbf{r})\right|\varphi_{\mathbf{k}}(\mathbf{r})\right\rangle
\end{equation}
is the matrix element of the scattering potential,
\begin{equation}\label{Af}
f_{\mathbf{k}^\prime\mathbf{k}}=\frac{2E_\omega
ek}{m\omega^2}\sin\left(\frac{\theta}{2}\right),
\end{equation}
and
$\theta=\varphi-\varphi^\prime=(\widehat{\mathbf{k},\mathbf{k}^\prime})$
is the scattering angle. Let us apply the Jacobi-Anger expansion,
$$e^{iz\sin\gamma}=\sum_{n=-\infty}^{\infty}J_n(z)e^{in\gamma},$$
in order to rewrite Eq.~(\ref{Aak}) as
\begin{eqnarray}\label{Awk}
|a_{\mathbf{k}^\prime}(t)|^2&=&\frac{\left|U_{\mathbf{k}^\prime\mathbf{k}}\right|^2}{\hbar^2}
\Bigg|\sum_{n=-\infty}^{\infty}{J_n\left(f_{\mathbf{k}^\prime\mathbf{k}}\right)}\,
e^{i(\varepsilon_{k^\prime}-\varepsilon_k+n\hbar\omega)t/{2\hbar}}\nonumber\\
&\times&e^{in(\varphi+\varphi^\prime)/2}
\int\limits_{-t/2}^{t/2}e^{i(\varepsilon_{k^\prime}-\varepsilon_k
+n\hbar\omega)t^\prime/\hbar}dt^\prime\Bigg|^2,\nonumber\\
\end{eqnarray}
where $J_n(z)$ is the n-th order Bessel function of the first
kind. Since the integrals in Eq.~(\ref{Awk}) for long time $t$
turn into the delta-function
$$\delta(\varepsilon)=
\frac{1}{2\pi\hbar}\lim_{t\rightarrow\infty}
\int_{-t/2}^{t/2}e^{i\varepsilon t^\prime/\hbar}dt^\prime,$$ the
expression (\ref{Awk}) takes the form
\begin{equation}\label{wkk}
|a_{\mathbf{k}^\prime}(t)|^2=4\pi^2\left|U_{\mathbf{k}^\prime\mathbf{k}}\right|^2
\sum_{n=-\infty}^{\infty}J_n^2\left(f_{\mathbf{k}^\prime\mathbf{k}}\right)
\delta^2(\varepsilon_{k^\prime}-\varepsilon_k+n\hbar\omega).
\end{equation}
To transform the square delta-functions in Eq.~(\ref{wkk}), we can
apply the conventional procedure,
$$\delta^2(\varepsilon)=\delta(\varepsilon)\delta(0)
=\frac{\delta(\varepsilon)}{2\pi\hbar}\lim_{t\rightarrow\infty}
\int\limits_{-t/2}^{t/2}e^{i0\times
t^\prime/\hbar}dt^\prime=\frac{\delta(\varepsilon)t}{2\pi\hbar}.$$
Keeping in mind that $\varepsilon_k=\varepsilon_{k^\prime}$, the
probability of the electron scattering between the states
(\ref{Apsi}) with the wave vectors $\mathbf{k}$ and
$\mathbf{k}^\prime$ per unit time,
$w_{\mathbf{k}^\prime\mathbf{k}}=d|a_{\mathbf{k}^\prime}(t)|^2/d
t$, is given by
\begin{equation}\label{AW0}
w_{\mathbf{k}^\prime\mathbf{k}}=\frac{2\pi}{\hbar}J_0^2\left(f_{\mathbf{k}^\prime\mathbf{k}}\right)
\left|U_{\mathbf{k}^\prime\mathbf{k}}\right|^2
\delta(\varepsilon_{k^\prime}-\varepsilon_k).
\end{equation}
Since Eq.~(\ref{Af}) depends only on the electron energy
$\varepsilon_k=\hbar^2k^2/2m$ and the scattering angle $\theta$,
the probability (\ref{AW0}) describes the isotropic scattering
which can be tuned by the wave amplitude $E_\omega$ and the wave
frequency $\omega$.

\end{document}